\documentclass[3p,times,twocolumn]{elsarticle}

\usepackage{ecrc}

\volume{00}

\firstpage{1}

\journalname{Nuclear Physics B Proceedings Supplement}

\runauth{}

\jid{nuphbp}

\jnltitlelogo{Nuclear Physics B Proceedings Supplement}

\usepackage{amssymb}
\usepackage[figuresright]{rotating}

\usepackage{captcont}
\usepackage[final,		% override "draft" which means "do nothing"
	colorlinks,			% rather than outlining them in boxes
	linktocpage,		% link just the pages, not the full titles
	linkcolor=blue,		% override truly awful color choices
	citecolor=blue,		% (ditto)
	urlcolor=blue,		% (ditto)
	breaklinks=true,	% so that links wrap around to multiple lines if necessary
	]{hyperref}
\usepackage[all]{hypcap}	% to get figure links correct

\begin{document}

\begin{frontmatter}



\dochead{}

\title{Physics and astrophysics with gamma-ray telescopes}


\author{J.~Vandenbroucke, for the Fermi LAT collaboration}

\address{Kavli Institute for Particle Astrophysics and Cosmology, Department of Physics and SLAC National Accelerator Laboratory, Stanford University, Stanford, CA 94305, USA}

\begin{abstract}

 In the past few years gamma-ray astronomy has entered a golden age. A modern suite of telescopes is now scanning the sky over both hemispheres and over six orders of magnitude in energy. At $\sim$TeV energies, only a handful of sources were known a decade ago, but the current generation of ground-based imaging atmospheric Cherenkov telescopes (H.E.S.S., MAGIC, and VERITAS) has increased this number to nearly one hundred. With a large field of view and duty cycle, the Tibet and Milagro air shower detectors have demonstrated the promise of the direct particle detection technique for TeV gamma rays. At $\sim$GeV energies, the Fermi Gamma-ray Space Telescope has increased the number of known sources by nearly an order of magnitude in its first year of operation. New classes of sources that were previously theorized to be gamma-ray emitters have now been confirmed observationally. Moreover, there have been surprise discoveries of GeV gamma-ray emission from source classes for which no theory predicted it was possible. In addition to elucidating the processes of high-energy astrophysics, gamma-ray telescopes are making essential contributions to fundamental physics topics including quantum gravity, gravitational waves, and dark matter. I summarize the current census of astrophysical gamma-ray sources, highlight some recent discoveries relevant to fundamental physics, and describe the synergetic connections between gamma-ray and neutrino astronomy. This is a brief overview intended in particular for particle physicists and neutrino astronomers, based on a presentation at the Neutrino 2010 conference in Athens, Greece. I focus in particular on results from Fermi (which was launched soon after Neutrino 2008), and conclude with a description of the next generation of instruments, namely HAWC and the Cherenkov Telescope Array.

\end{abstract}


\end{frontmatter}

% \linenumbers


\begin{figure}[t!]
\centering
\noindent\includegraphics[width = 0.5\textwidth]{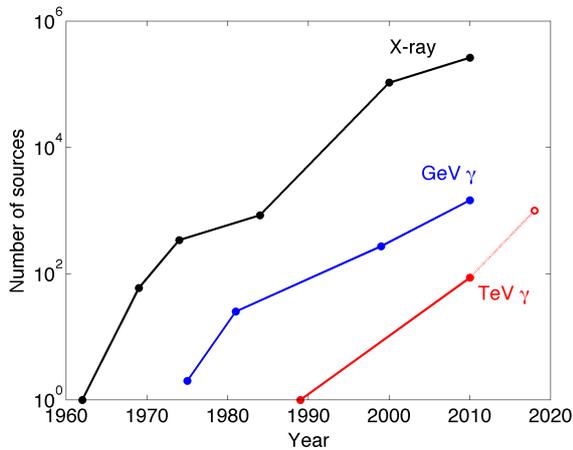}
\caption{``Kifune'' plot summarizing progress in high-energy astrophysics over the past five decades in terms of source counts in three energy bands.  The plot has been updated to include the most recent catalogs from 2010: the 2XMMi-DR3 list of 262,902 X-ray sources~\cite{2XMMi-DR3}, the 1FGL list of 1451 GeV gamma-ray sources~\cite{1FGL}, and the TeVCat list of 86 TeV gamma-ray sources~\cite{TeVCat}.  It also includes a projection based on simulations of CTA performance and TeV population studies, which predicts that $\sim$1000 TeV sources will be detected by CTA~\cite{CTASources}.}
\label{kifune}
\end{figure}

\section{Experimental techniques and the current generation of telescopes}

%ARGO-YBJ
%Tibet AS gamma

Charged cosmic rays at the low-energy end of the spectrum are best studied by balloon- and space- borne instruments, because their penetration length into the atmosphere is small but their flux is large enough to be detected by a reasonably small payload.  At the high-energy end the opposite is true: the flux of particles is smaller than can be measured by a satellite, but the interactions develop into extensive air showers which can be detected by various techniques and used to estimate the properties of the primary cosmic ray from the ground.  Similar principles apply to gamma rays as to charged cosmic rays: Below $\sim$100~GeV, typical sources have fluxes large enough to be detected directly by particle detectors that can fit reasonably on a satellite and fly above the atmosphere; above $\sim$100~GeV the primary gamma rays produce air showers that can be detected by ground-based instruments and used to determine the energy and direction of the primary gamma ray.

Two gamma-ray satellites optimized for the GeV~energy range are currently operating: AGILE (Astro-rivelatore Gamma a Immagini LEggero) and the Fermi Gamma-ray Space Telescope.  AGILE~\cite{AGILE} was launched on April 23, 2007 and features both an 18-60~keV X-ray monitor and a gamma-ray pair-production telescope with a peak effective area of $\sim$600~cm$^2$ up to $\sim$10~GeV.  Fermi was launched on June 11, 2008 and has two instruments: the Gamma-ray Burst Monitor (GBM) and the Large Area Telescope (LAT~\cite{LAT}).  The GBM is sensitive to X-rays and gamma rays, from 8~keV to 40~MeV.  The LAT is sensitive to gamma rays from 20~MeV to more than 300~GeV, with a peak effective area of $\sim$8000~cm$^2$ above a few GeV.  The LAT is a pair-production telescope with three components: a silicon tracker to determine the direction of each event, a calorimeter featuring thallium-doped cesium iodide scintillating crystals to determine the energy of each event, and an anti-coincidence detector surrounding the tracker to distinguish gamma-ray events from the much more numerous charged-particle events.  Pair conversion occurs predominantly in dense tungsten foils that alternate with silicon strip detection planes in the tracker.  The LAT has a large (2.4~sr) field of view.  Since commissioning, Fermi has operated primarily in sky-survey mode, alternating its viewing direction every $\sim$1.5~hr orbit to achieve nearly uniform coverage of the full sky every two orbits.

%Exceptions to the survey mode occur when the spacecraft autonomously slews to follow a gamma-ray burst for several hours after it is detected and ground-directed Target of Opportunities when exceptional activity from a single source warrants pointing at it continuously for an extended time up to several days.

At energies above $\sim$100~GeV, typical source fluxes are too low to be measured efficiently by a detector that can fit on a satellite or ballon.  Nevertheless, the innovative imaging atmospheric Cherenkov telescope (IACT) technique has enabled ground-based detectors to achieve large effective areas in the TeV range.  This technique, established in 1989 with the detection of the Crab Nebula at the Whipple 10~m telescope~\cite{Weekes}, uses telescopes to detect the Cherenkov light produced by charged particles in air showers.  Because the Cherenkov angle in air is small ($\sim$1$^\circ$), the Cherenkov light comes from the same direction as the primary gamma ray and therefore an IACT must be pointed at a source to detect it.  Current IACTs feature segmented single-dish telescopes that reflect the Cherenkov light to a camera composed of photomultiplier tubes (PMTs) and fast readout electronics, and have fields of view a few degrees across.  Fast cameras are necessary to discriminate the few-nanosecond Cherenkov pulse from the steady night-sky background light.  As at $\sim$GeV energies, charged cosmic-ray particles are significantly more numerous than gamma rays at $\sim$TeV energies and constitute a background that must be reduced.  Shape characteristics of the air shower are used to distinguish these cosmic-ray-induced hadronic showers from the gamma-ray-induced electromagnetic showers.  The residual background from hadronic showers is estimated and subtracted by comparing on-source and off-source regions.  IACTs can only operate during nights with low moonlight and good weather and therefore typically accrue $\sim$1000~hours of good observing time per year, corresponding to a duty cycle of $\sim$10\%.
% Note the "wobble" method is somewhat different from the on-off method.

The current generation of IACTs includes H.E.S.S. (the High Energy Stereoscopic System) in Namibia (23$^\circ$~S latitude, 1800~m altitude); MAGIC (Major Atmospheric Gamma-ray Imaging Cherenkov Telescope) on La Palma in the Canary Islands (29$^\circ$~N latitude, 2200~m altitude), and VERITAS (Very Energetic Radiation Imaging Telescope Array System) in Arizona (32$^\circ$~N latitude, 1300~m altitude).  H.E.S.S. consists of 4 telescopes each with a diameter of 13~m, operating since 2004.  A fifth telescope with a diameter of 28~m (H.E.S.S. II) is under construction and will lower the energy threshold of H.E.S.S from 100~GeV to 30~GeV.  MAGIC consists of 2 telescopes (one operating since 2004 and the other since 2009), each with a 17~m diameter.  VERITAS consists of 4 telescopes, each with a 12~m diameter, operating since 2007.

% CANGAROO III: 4 telescopes, 10 m diameter, Woomera, South Australia.
% Add latitudes.
In addition to imaging atmospheric Cherenkov telescopes, it is possible to detect gamma-ray-induced air showers using particle detectors at ground level.  Although such detectors have a high energy threshold and a high background contamination rate, they have much larger fields of view and duty cycles than IACTs and can provide continuous monitoring of a large fraction of the sky, with relatively uniform exposure well-suited for unbiased surveys.  The Tibet~\cite{Tibet} and Milagro~\cite{Milagro} air shower detectors have pioneered this technique.
% Cross-correlation of the Fermi and Milagro data sets recently produced a list of Galactic sources with evidence of TeV emission~\cite{Milagro}.

% ARGO-YBJ
% Tibeg AS-gamma

%At GeV energies, the p-gamma cross section.

%Note that the resolution with which the telescopes can determine the position of a particular source increases with its brightness.

%MILAGRO

%\begin{table*}[hbt]
\begin{table*}[]
\caption{Number of known GeV and TeV gamma-ray sources by type, as of November 2010.  AGN, the most abundant source class, are divided into sub-classes.  GeV sources are from the one-year Fermi LAT catalog (1FGL~\cite{1FGL}).  TeV sources are from the ``Default Catalog'' of TeVCat~\cite{TeVCat}, which does not include some newly announced sources.  The two non-blazar AGNs detected at TeV are Fanaroff-Riley Type I radio galaxies.  For Fermi we only include those sources listed in the first-year catalog; more sources have been detected in dedicated analyses.  Some sources have been firmly identified, and some are only positionally associated, with counterparts in other wave bands; see~\cite{1FGL}.  For the GeV sources we follow Table 6 of ~\cite{1FGL}.  The total number of point sources in 1FGL is 1451, while the sum of the rows in the table is 1454, because three objects (Crab, Vela, and MSH~15-52) are included in both the pulsars and the pulsar wind nebulae row.  Five sources are associated with the Large Magellanic Cloud, so the number of unique sources is $1451-4=1447$.  There are many unidentified sources at both GeV and TeV energies.  Work is underway to identify them with multi-wavelength observations and to classify them according to their gamma-ray spectral and variability properties.}
% Among the 23 BL Lac blazars detected by TeV telescopes, 19 are high-frequency-peaked (HBL), 2 are intermediate-frequency-peaked (IBL), and 2 are low-frequency-peaked (LBL)
% I exclude "other" sources from TeVCat.  These are sources which are somewhat conflicting/contraversial.
% Pulsars: 39 radio loud + 24 radio quiet
\label{census}
 \begin{center}
    \begin{tabular}{|l|c|c|} \hline
      {\bf Source class} & {\bf GeV sources} & {\bf TeV sources} \\ \hline
      Unidentified & 630 & 26 \\ \hline % TeV: 24 unidentified + 2 other = 26
      AGN: BL Lac blazars & 295 & 23 \\ \hline
      AGN: Flat-spectrum radio quasar (FSRQ) blazars & 278 & 1 \\ \hline
      AGN: non-blazar & 28 & 2 \\ \hline
      AGN: uncertain type & 92 & 0 \\ \hline
      Pulsars & 63 & 0 \\ \hline
      Shell-type supernova remnants & 44 & 10 \\ \hline
      Pulsar Wind Nebulae (PWN) & 5 & 18 \\ \hline
      Globular clusters & 8 & 0 \\ \hline
      X-ray binaries & 3 & 3 \\ \hline
      Starburst galaxies & 2 & 2 \\ \hline % M82 and NGC 253
      Normal galaxies & 2 & 0 \\ \hline
      Wolf-Rayet stars & 0 & 1 \\ \hline	% Westerlund 2
%      Dark & 0 & 1 \\ \hline
%      Other & x & 2 \\ \hline
      \bf{Total} & \bf{1451} & \bf{86} \\ \hline
     \end{tabular}
  \end{center}
\end{table*}

\section{A census of sources}

Figure~\ref{kifune} shows an updated version of the ``Kifune plot'' (first produced by T.~Kifune), showing the number of known X-ray, GeV gamma-ray, and TeV gamma-ray sources as a function of time.  Here we use ``GeV sources'' to denote sources detected by space-based pair-production telescopes, and ``TeV sources'' to denote sources detected by ground-based air-shower detectors.
% Pioneering work of EGRET, Whipple, SAS, OSO, HEGRA, ...

\begin{figure}[t!]
\centering
\noindent\includegraphics[width = 0.5\textwidth]{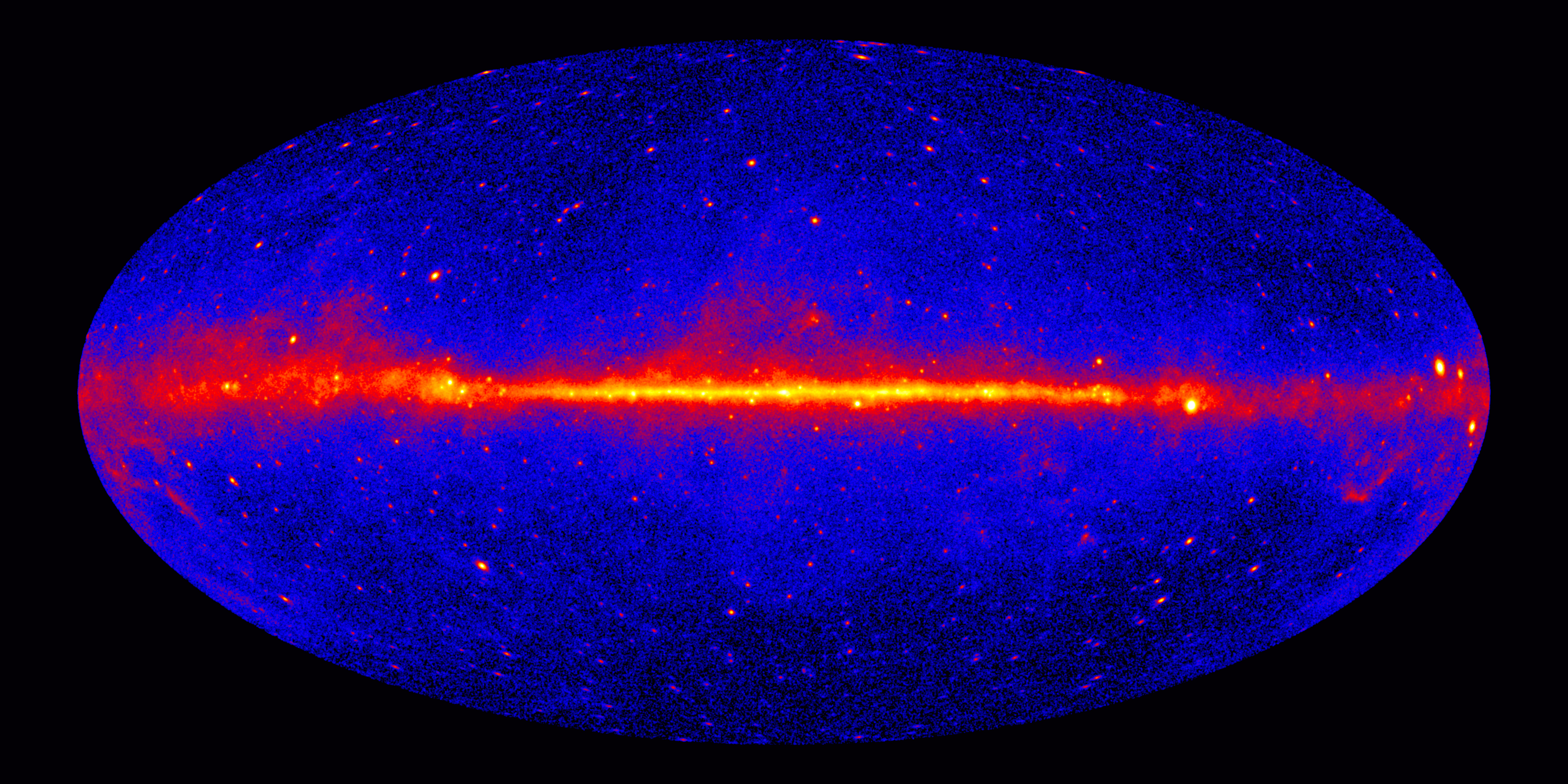}
\caption{All-sky gamma-ray map from two years of Fermi LAT data, in Galactic coordinates and Aitoff projection.  The color scale indicates the number of photons with energy greater than 1~GeV detected per pixel in the first twenty-four months of LAT operation.  Diffuse Galactic emission is evident as a bright red band across the Galactic plane.  This diffuse emission is produced by cosmic rays colliding with diffuse gas in the Milky Way, and by inverse Compton scattering of background starlight by energetic electrons.  Bright point sources are visible both within the Galactic disk (predominantly pulsars, pulsar wind nebulae, and supernova remnants) and outside of it (predominantly AGNs, along with some millisecond pulsars).  Many of the point sources are still unidentified.}
\label{skymap}
\end{figure}

Table~\ref{census} lists the current number of gamma-ray sources in the GeV and TeV range, separated according to source class.  While Fermi's predecessor EGRET detected 271 sources in nine years of operation (five years of which were used to produce the third and final EGRET catalog), the Fermi LAT detected 1451 sources in its first year of operation.  Moreover, the only steady source classes detected by EGRET were active galactic nuclei (AGN), pulsars, and one normal galaxy (the Large Magellanic Cloud).  Fermi has added pulsar wind nebulae, shell-type supernova remnants, X-ray binaries, starburst galaxies, and globular clusters to this list.  24 of the 63 pulsars Fermi has detected were discovered for the first time in gamma rays and had not been previously detected in radio waves; some of these have since been detected in the radio band but others have not despite deep observations.  Fermi reached the same sensitivity in four days that EGRET reached in one year.  The two-year Fermi sky map is shown in Figure~\ref{skymap}.

\section{Recent highlights from GeV to TeV}

Within several weeks of one another, H.E.S.S., VERITAS, and Fermi announced the first detection of both GeV and TeV gamma-ray emission from starburst galaxies~\cite{starburstHESS,starburstVERITAS,starburstFermi}.  While gamma-ray emission from normal galaxies in the Local Group, such as the Large Magellanic Cloud, as well as from distant galaxies with active nuclei, had already been established, starburst galaxies had been predicted to fill the gap between these extremes, with intermediate distances and intermediate theorized gamma-ray luminosities.  Starburst galaxies are galaxies with large rates of star formation occurring in regions rich in molecular gas.  The large rate of massive star production results in a correspondingly large rate of supernovae, whose remnants could be accelerating cosmic rays which then collide with the ambient gas to produce gamma rays.  The discovery of gamma rays from starburst galaxies supports this basic picture.

Fermi has increased the number of gamma-ray-emitting pulsars by an order of magnitude from the 6 detected by EGRET.  In particular, 18 millisecond pulsars were detected in the first few months of Fermi operation~\cite{msp}.  Millisecond pulsars are neutron stars spinning very fast, hundreds of times per second.  Prior to the launch of Fermi, 60 of them were known (not including those in globular clusters), all discovered by radio telescopes.  By monitoring the timing of a network of such millisecond pulsars, the passage of a gravitational wave through the network could be detected~\cite{pta}.  Millisecond pulsars, as opposed to standard young pulsars with much larger ($\sim$30~ms to $\sim$10~s) rotation periods, are necessary for gravitational wave detection because they are significantly more stable than young pulsars and can therefore serve as good reference clocks.  Such ``pulsar timing arrays'' are sensitive to very low frequency ($\sim$nHz) waves from astrophysical sources such as supermassive black hole mergers and are complementary to ground-based ($\sim$kHz) and even space-based ($\sim$mHz) gravitational wave interferometers that are sensitive to higher frequencies.

% GBM bursts: http://heasarc.gsfc.nasa.gov/cgi-bin/W3Browse/w3catindex.pl#FERMI
% LAT bursts:	http://fermi.gsfc.nasa.gov/ssc/resources/observations/grbs/grb_table/
% 			https://www-glast.stanford.edu/cgi-bin/pub_rapid#gcn
% Detected GRBs:
%	080825C
%	080916C
%	081024B
%	081215A
%	081224A
%	091217
%	090323
%	090328
%	090510
%	090626
%	090902B
%	090926
%	091003A
%	NOT 091010 (upper limits)
%	091031
%	100116A
%	100225A
%	100325A
%	100414A
%	100707A
%	100724B
%	100826A
%	101014A
As of November 2010, the Gamma-ray Burst Monitor on Fermi has detected 594 bursts, 22 of which have also been detected in high-energy gamma rays by the LAT.  One particular burst at a redshift of 0.903 was detected over a very broad energy range from 8~keV to 31~GeV, spanning a factor of $4 \times 10^6$ in energy.  Despite this broad energy range and the fact that the burst of photons was traveling for seven billion years before reaching Fermi, the rising edge time of the burst varied by less than 1~second over the broad energy range.  Many theories of quantum gravity predict Lorentz invariance violation (LIV) that should be detected in such a burst as an energy-dependent variation in the rising edge time of the burst.  Detection of such a delay could be due to LIV or to intrinsic emission time variation; the non-observation of such a delay in this burst allows Fermi to constrain the Lorentz invariance length scale to be less than the Planck length divided by 1.2, at 99\% confidence~\cite{fermiQuantumGravity}.  This result is now the strongest constraint on LIV by several orders of magnitude.
%The previous strongest constraint was achieved by applying the same idea to a fast-varying blazar detected by H.E.S.S.~\cite{hessQuantumGravity}.

In addition to constraining theories of quantum gravity, gamma-ray telescopes can test general relativity.  Some distant blazars are strongly gravitationally lensed, such that two images are resolved by optical telescopes.  If the blazar flares, a time lag (in some cases as large as several weeks) can be detected between the flare time in the two images, due to the different propagation time through the two gravitationally deflected light paths.  Gamma-ray telescopes cannot independently resolve such double images.  However, if a flare is detected from such a system at both optical and gamma-ray wavelengths, comparison of the arrival time of the two flares in the two wavelengths can test whether photons with energy that differs by eight orders of magnitude are both deflected in the same way, as predicted by general relativity.  Such a system was recently detected flaring by Fermi~\cite{lensBlazar}.
%The presence of two lens images in this system, corresponding to two different deflected paths, allows us to separate intrinsic delays between the optical and gamma-ray emission by the source from possible differences in propagation of gamma rays with respect to optical photons, by comparing the time difference between the two optical flares with the time difference between the two gamma-ray flares.

There has been significant excitement in the past two years over surprising results from the ATIC and PAMELA instruments: ATIC detected a sharp bump in the combined electron plus positron spectrum at several hundred GeV, and PAMELA detected an increase in the positron fraction between 10 and 100 GeV.  Although both Fermi and H.E.S.S. were designed to detect gamma rays, they have also measured the primary cosmic-ray electron plus positron spectrum better than any other instrument in their respective energy ranges.  The cosmic ray electron plus positron population could include contributions from diffuse secondary production (by collisions of cosmic rays with diffuse gas), from discrete primary sources such as pulsars, and from dark matter annihilation.  Although the spectrum measured by Fermi~\cite{fermiCRE} is somewhat harder than that predicted by GALPROP~\cite{GALPROP} models of diffuse secondary production, it is not consistent with the sharp bump detected by the ATIC instrument.  The spectrum softens significantly at TeV energies as measured by H.E.S.S.~\cite{hessCRE}  In addition to these measurements of the combined electron plus positron spectrum, Fermi can cross-check the PAMELA measurement of the positron fraction by using the cosmic ray East-West effect: by selecting particular directions in the Earth's magnetic field, pure electron or positron samples can be selected and used to measure the absolute positron-only spectrum.
%The opposite charge is forbidden to arrive from these directions because its path through the geomagnetic field from space would have been blocked by the Earth.  This, along with results from the AMS detector to be installed on the space station in 2011, will extend and cross-check the surprising measurement from PAMELA that the positron fraction increases significantly in the 10-100~GeV range.

Both GeV and TeV gamma-ray telescopes can contribute significantly to constraining models of dark matter by searching for gamma rays from dark matter annihilation.  Such signals could be detected from galaxy clusters, dwarf satellite galaxies, the Galactic center, the Milky Way halo, or as a diffuse signal from many individual galaxies.  Ultra-faint dwarf spheroidal galaxies are a particularly promising target.  These are very small galaxies orbiting as satellites of the Milky Way, with only a few hundred stars but with as much as 100 times as much mass in dark matter.  They are thus very clean targets for gamma-ray telescopes because there is expected to be minimal astrophysical background.  Searches for gamma rays from dwarf spheroidal satellites have been performed with Whipple, VERITAS, H.E.S.S., MAGIC, and Fermi (see~\cite{fermiDwarf} and references therein), and their non-detection has provided some of the most stringent constraints on WIMP models.  Fewer than 20 ultra-faint dwarf satellites are known~\cite{dwarf}.  They have been detected in optical surveys such as the Sloan Digital Sky Survey, which have covered a limited fraction of the sky.  It is expected that many more dwarf satellites will be discovered with upcoming optical surveys such as the Dark Energy Survey and the Large Synoptic Survey Telescope (LSST) and could be used to provide even stronger gamma-ray constraints on WIMP models.  It is even possible that some of the existing unidentified gamma-ray sources are dwarf galaxies for which an optical counterpart has not yet been identified; spectral analysis of unidentified Fermi sources is underway to determine such candidates.

% Recently approved 24-month sky map by Seth
% from https://confluence.slac.stanford.edu/display/SCIGRPS/2010/10/19/24-month+all-sky+image

\section{Gamma-neutrino connections}

%Ahlers et al GZK neutrinos

% pair-production cross section (gamma on H or He): ~20 mb from ~0.5 to ~4 GeV
% http://www.sciencedirect.com/science?_ob=ArticleURL&_udi=B6TVC-4718HV8-39&_user=3828959&_coverDate=08/01/1972&_rdoc=1&_fmt=high&_orig=search&_origin=search&_sort=d&_docanchor=&view=c&_searchStrId=1563849129&_rerunOrigin=google&_acct=C000012078&_version=1&_urlVersion=0&_userid=3828959&md5=7638d8b6b384572abda2314d10d05fc7&searchtype=a
% Thomson cross section: 0.7 barn = 700 mbarn
% particle data book: pair production in nuclear field (~GeV to ~TeV) has a cross section of ~300 mb on carbon or ~30 barn on lead or 
Gamma-ray and neutrino astronomy share two of their fundamental goals: understanding the sources and mechanisms of cosmic ray acceleration, and elucidating the nature and distribution of dark matter in the universe.  Due to their much larger interaction cross sections, gamma rays are significantly easier to detect than neutrinos, and gamma-ray astronomy has made significantly more progress to date.  The number of known astrophysical neutrino sources remains two: the Sun and Supernova 1987A.  The existence of cosmic-rays, however, guarantees the existence of more astrophysical neutrino sources.

The leading Galactic candidates for cosmic-ray sources are supernova remnants.  The leading extragalactic candidates are AGN, gamma-ray bursts (GRBs), and galaxy clusters.  While gamma rays can be produced by hadronic ($\pi^0$ decay) or electromagnetic (synchrotron, inverse Compton, or bremsstrahlung) processes, neutrinos can only be produced by hadronic processes ($\pi^\pm$ decay).  Roughly speaking, the detection of gamma rays from a source is therefore necessary but not sufficient for determining the source to be a cosmic ray accelerator.  While detailed modeling of gamma-ray morphology and spectra has in some cases produced a preference for hadronic models (particularly for certain supernova remnants), it is generally possible to fit the gamma-ray data with either electromagnetic or hadronic models.  No single source, either Galactic or extragalactic, has been conclusively proven to be a hadronic cosmic ray accelerator.

While determining the production mechanism of gamma rays is challenging, the detection of neutrinos from a cosmic ray accelerator candidate would provide unambiguous ``smoking gun'' evidence that it is in fact accelerating hadrons.  Conversely, non-detection of neutrinos from particular sources with neutrino telescopes such as IceCube and ANTARES will be sufficient to rule them out as hadronic accelerators.  Gamma-ray telescopes are essential to neutrino astronomy because they can tell us where and when to look for neutrinos.  The specific characteristics (light curves and energy spectra) of each gamma-ray source can be used to make specific predictions for neutrino signals (under the hypothesis of hadronic gamma-ray production) and optimize searches for them.

% Swift BAT fov is 2 str: http://www.swift.ac.uk/instruments.shtml
% Fermi GBM fov is whole sky not occluded by earth (more than 2pi): http://en.wikipedia.org/wiki/Fermi_Gamma-ray_Space_Telescope
Gamma-ray light curves can be used to search for neutrinos during blazar flares, during particular phases in periodic sources such as X-ray binaries, and during the brief transient emission of GRBs.  The prospects for neutrino searches in IceCube data using Fermi-detected GRBs are particularly promising.  The wide field of view of the GBM (it can detect GRBs in the entire sky not occluded by the Earth) allows it to detect a high rate of nearby, bright bursts, which are the bursts most promising for neutrino detection.  An IceCube search for neutrino emission during GRBs detected in 2008, when only half of IceCube was complete and Fermi was operating for only a few months, has already yielded upper limits at the level of model predictions~\cite{Kappes}.  Subsequent years of IceCube data, with more of the instrument complete and Fermi operating full time, can confirm or rule out GRBs as cosmic ray sources.

In addition to gamma-ray timing, gamma-ray spectra can be used to predict neutrino signals and optimize the search for them.  In~\cite{Neronov09}, for example, Neronov and Ribordy fit hadronic models to individual blazar spectra from Fermi and IACT data and use them to predict neutrino spectra and event rates at IceCube.  They identify 13 Northern-hemisphere (in the field of view of IceCube) blazars from which significant ($\sim$5$\sigma$) neutrino detection is predicted in one year of operation of the full IceCube detector, if the GeV gamma-ray emission is dominated by pion decay.  These predictions are for steady-state quiescent emission; detection of a smaller number of neutrinos from the direction of a blazar could be significant if coincident with flaring activity of the blazar.  Among the two blazar classes, BL Lacs are more promising candidate neutrino sources than flat-spectrum radio quasars because they produce more energy at TeV energies and are therefore well matched to the $\sim$100~GeV energy threshold of neutrino telescopes.
% With construction to be completed at the end of 2010, IceCube will confirm or reject models of cosmic-ray acceleration in AGNs in the next couple years.

Gamma-ray and neutrino telescopes are also complementary in the study of dark matter.  In addition to direct searches for dark matter and searches for dark matter produced at man-made accelerators, gamma-ray and neutrino telescopes can study the annihilation of dark matter particles \emph{in situ}.  While neutrino telescopes can search for dark matter annihilation in many of the same sources as gamma-ray telescopes, the predicted rates are in general not competitive with the gamma-ray searches.  However, neutrino telescopes can also search for dark matter annihilation by WIMPs trapped in the gravitational wells of the Earth and the Sun.  While gamma rays produced from such annihilations would not escape the dense body, searches for neutrinos from these WIMPs have provided some of the most stringent constraints on dark matter models~\cite{sunWIMPs}.

%As for cosmic ray accelerator candidates however, gamma-ray telescopes suffer to some extent from an abundance of wealth.  While there is 

%neutrinos essential for disentangling...

%Gamma-ray bursts and neutrinos and the benefit of Fermi.

\section{The next generation}

%quantum gravity

%time domain: periods (binaries), flares (pwn, blazars) and transients (grbs)

%Improvements to Fermi... pi0 decay bump.
%CTA design studies: cite simulation studies and design document.

%positron fraction

%difficulty of diffuse

%difficulty of angular resolution

%anisotropy

%new grb sky map?

%Milagro sources confirm Fermi... HAWC potential.

Following the success of the current generation of IACTs, the Cherenkov Telescope Array (CTA~\cite{CTADesign}) is being designed to achieve an order of magnitude larger sensitivity at 1~TeV.  CTA will benefit from experience of the current generation of IACTs and scale the technology up to an array of dozens of individual telescopes.  In addition to a much larger number of telescopes, CTA will feature an improved field of view, angular resolution, and hadronic background rejection capability.  In addition to increasing the statistics of known source classes to enable population studies that are currently impossible, CTA will likely discover entire new classes of sources that produce TeV emission, as Fermi has done for GeV emission.  Finally, for the sources that are already known to produce TeV emission, the improved sensitivity and angular resolution of CTA will enable studies of light curves, spectra, and morphology for sources that are now known to emit TeV gamma-rays but with insufficient statistics to perform the binning necessary for such studies.  Such detailed analysis is necessary to determine \emph{how} these sources produce TeV gamma-rays, now that we know \emph{that} they do.

% Construction of HAWC should be complete in 2014.  Each tank will hold three 8" PMTs.  Single gamma-ray resolution will be 0.1 deg.  Milagro's altitude was 2630~m.  Seven tanks have been installed as of Fall 2010.  One has been filled with water.  The seven tanks will be in operation in early 2011.  By fall 2011, 30 tanks expected to be installed.
Following the success of Milagro, the High Altitude Water Cherenkov (HAWC) detector is under construction to monitor a large fraction ($\sim$2~sr) of the sky continuously with high sensitivity ($\sim$15 times larger than Milagro) from 100~GeV to 100~TeV~\cite{Taboada}.  HAWC is under construction at Sierra Negra, Mexico, a very high altitude (4100~m) site, and re-uses many of the components of Milagro.  While Milagro consisted primarily of a single large water pool instrumented with PMTs, HAWC will feature 300 discrete large cylindrical water tanks (7.3~m diameter by 4.5~m deep).  This segmentation allows better determination of the air shower shape, which enables good discrimination between smooth gamma-ray-induced electromagnetic showers and clumpy cosmic-ray-induced hadronic air showers.  The large field of view, high (nearly 100\%) duty cycle, and good sensitivity of HAWC will enable new discoveries from directions that IACT instruments do not select to observe.  In particular, HAWC's large field of view and duty cycle are well suited to observe GRBs, particularly in light of the high-energy component of GRB emission established by Fermi~\cite{Taboada}.

It is worth emphasizing that the gamma-ray sky is both highly variable and highly anisotropic.  We can therefore expect to continue making new discoveries at a rate that scales almost directly with how long we observe and in how many directions we observe.  The fact that Fermi observes the whole sky every few hours has enabled significant discoveries through transient events.  In addition to GRBs, recent discoveries enabled by transients include the discovery of new blazars that were unidentified sources in their quiescent state but became identifiable due to a brief flare~\cite{blazarFlares}; the first detection of GeV gamma-rays from a nova~\cite{nova}, a source class which had never been theorized capable of producing such high-energy emission; and a fast and bright flare from the Crab nebula, which provides evidence that it is accelerating electrons to $10^{15}$~eV, the highest energy particles ever associated with an individual source~\cite{crab}.  The diversity of these transient events discovered in the first two years of Fermi operation indicates that, for the foreseeable future, similar discoveries will accrue in proportion to the time we spend observing the sky.  Indeed, such ``time-domain astronomy'' will continue to yield important discoveries across the electromagnetic spectrum, enabled by sensitive, wide-field instruments including Fermi, HAWC, and the LSST.

IACTs have a limited field of view and new discoveries can therefore be expected to continue as long as they continue to perform deep observations in more directions.  Valuable guidance on where to point the Cherenkov Telescope Array will be provided by Fermi, HAWC, and the current generation of IACTs.  Moreover, there is great synergy among these instruments because of their complementary energy and direction coverage.  By extrapolating spectra to predict how bright Fermi sources are in the TeV range, Fermi discoveries have led to Milagro and IACT discoveries.  Simultaneous operation of Fermi, HAWC, and the Cherenkov Telescope Array will allow such cross-fertilization to continue with unprecedented sensitivity over the next decade.  Finally, Fermi and the IACTs are providing constraints on hadronic acceleration models in astrophysical particle accelerators, which are being used to optimize the search for neutrinos from  such sources to confirm or reject that they are cosmic ray sources.  The synergies among gamma-ray and neutrino telescopes could pay off with significant discoveries about both the origin of cosmic rays and the nature of dark matter over the next decade.

\section{Acknowledgments}

% Taken from https://confluence.slac.stanford.edu/display/SCIGRPS/Acknowledgements+for+Proceedings
% and fixed the last sentence
The $Fermi$ LAT Collaboration acknowledges support from a number of agencies and institutes for both development and the operation of the LAT as well as scientific data analysis. These include NASA and DOE in the United States, CEA/Irfu and IN2P3/CNRS in France, ASI and INFN in Italy, MEXT, KEK, and JAXA in Japan, and the K.~A.~Wallenberg Foundation, the Swedish Research Council and the National Space Board in Sweden. Additional support from INAF in Italy and CNES in France for science analysis during the operations phase is also gratefully acknowledged.  J. Vandenbroucke is supported by a Kavli Fellowship from the Kavli Foundation.  I would like to thank the organizers of Neutrino 2010 for the invitation to present this overview.

%\label{}



%% References


\bibliographystyle{elsarticle-num}
\bibliography{vandenbroucke_neutrino_2010_proc}






% \end{thebibliography}

\end{document}